\documentclass[aps,prl,twocolumn,superscriptaddress,showpacs,
nofootinbib,nobalancelastpage,nobibnotes]{revtex4}
\usepackage[dvips]{epsfig}
\usepackage{amsmath}

\begin{document}

\title{Unified Graphical Summary of Neutrino Mixing Parameters}

\author{\large Olga Mena and Stephen Parke}

\affiliation{Theoretical Physics Department,\\ Fermi National Accelerator
Laboratory,\\ P.O.Box 500, Batavia, Illinois 60510, U.S.A.\\
Electronic addresses: omena@fnal.gov and parke@fnal.gov}

\date{Dec 10, 2003}

\begin{abstract}
The neutrino mixing parameters are presented in a number of different
ways by the various experiments, e.g.
SuperKamiokande, K2K, SNO, KamLAND and Chooz and also by the 
Particle Data Group.
In this paper, we argue that presenting the data in terms of 
$\sin^2 \theta$, where $\theta$ is the mixing angle appropriate for 
a given experiment has a direct physical interpretation.
For current atmospheric, solar and reactor
neutrino experiments, the $\sin^2 \theta$'s
are effectively the probability of finding a given flavor 
in a particular neutrino mass eigenstate.
The given flavor and particular mass eigenstate varies from experiment to
experiment, however, the use of $\sin^2 \theta$ provides a unified
picture of all the data.
Using this unified picture we present a graphical way to represent 
these neutrino mixing parameters
which includes the uncertainties. 
All of this is performed in the context of the present
experimental status of three neutrino oscillations. 

\end{abstract}

\pacs{14.60.Pq \hspace{2.25cm} hep-ph/0312131 
\hspace{2.25cm} FERMILAB-Pub-03/377-T}

\maketitle


Neutrino flavor transitions have been observed in atmospheric,
solar, reactor and accelerator experiments. 
Transitions for at least two different E/L's 
(neutrino energy divided by baseline) are seen.
To explain these transitions, extensions to the Standard Model 
of particle physics are required. 
The simplest and most widely accepted extension is to 
allow the neutrinos to have masses and mixings,
similar to the quark sector, then these flavor transitions can
be explained by neutrino oscillations.  

This picture of neutrino masses and mixings has recently come into 
sharper focus with the salt data presented by the SNO collaboration\cite{SNO}.
When combined with the KamLAND experiment\cite{KamLAND} and other
solar neutrino experiments\cite{SK_solar,solar_other} the range of allowed 
values for the solar mass squared difference, $\delta m^2_{sol}$, and 
the mixing angle, $\theta_{sol}$, are reported as
\begin{eqnarray}
6.6 \times 10^{-5} {\rm eV^2} < 
& \delta m^2_{sol} & 
< 8.7 \times 10^{-5} {\rm eV^2} \nonumber \\
0.33 < & \tan^2 \theta_{sol} & < 0.50 
\end{eqnarray}
at the 90 \% confidence level.  
Also maximal mixing, $\tan^2 \theta_{sol} =1$, has been ruled 
out at greater than 5 $\sigma$.
The solar data is consistent with $\nu_e \rightarrow \nu_\mu ~{\rm and/or}
~\nu_\tau$.

The atmospheric data from SuperKamiokande has changed only slight in 
the past year with a preliminary new analysis presented at EPS
conference\cite{SK_atm} and is consistent with the K2K 
long baseline experiment\cite{K2K}.
The ranged of allowed values for the atmospheric mass squared difference,
$\delta m^2_{atm}$ and the mixing angle, $\theta_{atm}$, are
reported as
\begin{eqnarray}
1.3 \times 10^{-3} {\rm eV^2} < 
& \delta m^2_{atm} & 
< 3.0 \times 10^{-3} {\rm eV^2} \nonumber \\
0.91 < & \sin^2 2\theta_{atm} & \leq 1  
\end{eqnarray}
at the 90 \% confidence level. 
The atmospheric data is consistent with
$\nu_\mu \rightarrow \nu_\tau$.  

The best constraint on the involvement of the $\nu_e$ at the atmospheric
$\delta m^2$ comes from the Chooz reactor experiment 
and this puts a limit on the mixing angle associated with 
these oscillations, $\theta_{chz}$, reported as
\begin{eqnarray}
0 \leq & \sin^2 2\theta_{chz} & < 0.1 
\end{eqnarray}
at the 90 \% confidence level at 
$\delta m^2_{atm} = 2.5 \times 10^{-3} {\rm eV^2}$. 
This constraint depends on the precise value of
$\delta m^2_{atm}$ with a stronger (weaker) constraint at higher
(lower) allowed values of $\delta m^2_{atm}$.

The only part of this picture which is still blurry is the
$\bar{\nu}_\mu \rightarrow \bar{\nu}_e$
flavor transitions reported by the LSND experiment\cite{LSND}
at large mass squared differences, 
$0.2 ~{\rm eV^2} < \delta m^2_{lsnd} < 1 ~{\rm eV^2} $.
Within the next couple of years the MiniBooNE experiment\cite{miniB}
will bring this part of the picture into focus.
If the flavor transitions claimed by the LSND experiment are confirmed
to be neutrino oscillations then a major revision of this 
picture maybe necessary\cite{recent_papers}. 
Given that the LSND observation is unconfirmed and doesn't
appear to fit easily into our current neutrino oscillation picture
we have decided to wait for confirmation by MiniBooNE before
trying to incorporate this result into the neutrino mixing 
parameter summary reported here.

\begin{figure*}
\epsfig{file=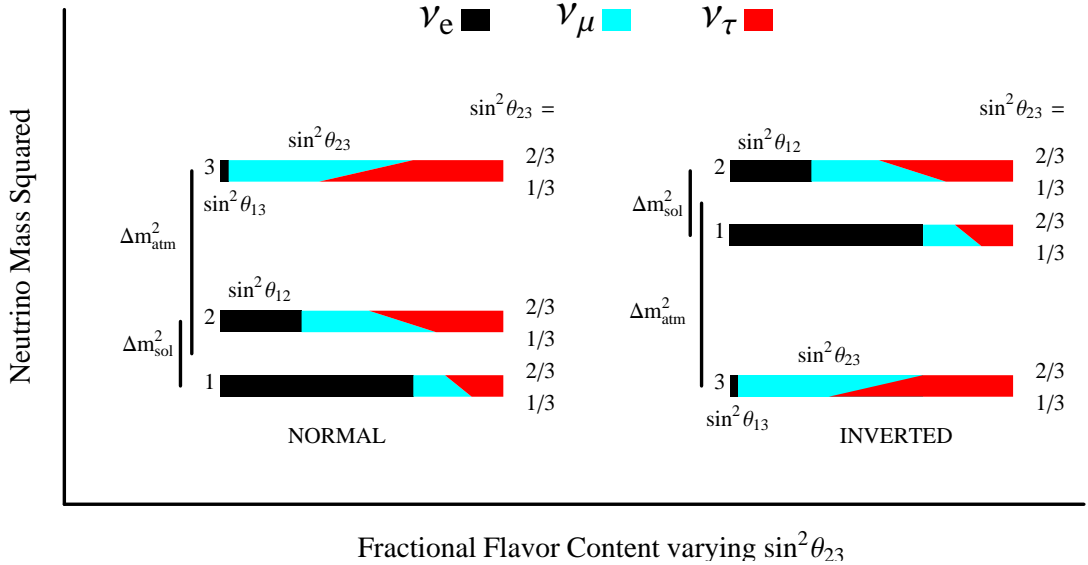, width=15cm} 
\caption{The range of probability of finding the $\alpha$-flavor in the i-th 
mass eigenstate as indicated for the two different mass hierarchies
as $\sin^2 \theta_{23}$ varies over its allowed 
range at the 90\% C.L. The bottom of the bars is for the minimum
allowed value of $\sin^2 \theta_{23} \approx 1/3$ and the top of the bars
is for the maximum value of   $\sin^2 \theta_{23} \approx 2/3$.
The other mixing parameters are held fixed:
$\sin^2 \theta_{12} = 0.30$, $\sin^2 \theta_{13} = 0.03$ and
$\delta = \frac{\pi}{2}$.}
\label{fig1}
\end{figure*}

The almost-standard way to incorporate the atmospheric, solar,
reactor and accelerator data into the three neutrino picture is to assign 
the small mass-squared
splitting associated with the solar 
$\delta m^2$ 
to the splitting between 
the mass eigenstates labeled 1 and 2.
Because of the effects of matter on solar neutrinos,
we already know that the mass eigenstate with
the larger electron neutrino component has the smaller mass.
We label this state 1 and the heavier state with the smaller electron neutrino
component state 2. 
Therefore the solar $\delta m^2$ can be identified\footnote{Our
convention is that $\delta m^2_{ji} \equiv m^2_j -m^2_i$.} as
\begin{equation}
\delta m^2_{21} \equiv m^2_2 - m^2_1 = \delta m^2_{sol} > 0.
\end{equation}
 
The large mass splitting associated with the atmospheric 
$\delta m^2$ is therefore the splitting between the mass eigenstate
labeled 3 
and the more closely spaced 1 and 2 mass eigenstates.
From the Chooz experiment we know that the 3-mass eigenstate
has a very small electron neutrino
component.
The sign of the splitting of this state from the solar doublet,
1 and 2, is unknown. Thus,
\begin{equation}
|\delta m^2_{32}|= \delta m^2_{atm} > 0.
\end{equation}
Therefore both mass eigenstates in the solar doublet, 1 and 2, 
could have smaller mass than
the 3 mass eigenstate (this is usually called the normal hierarchy)
or
a larger mass than the 3-mass eigenstate (the inverted
hierarchy).
The determination of the sign of this splitting, 
i.e the sign of $\delta m^2_{32}$, is one of the important
unknowns within the neutrino sector.

In the three neutrino scenario the mixing matrix which relates
the flavor states, $\alpha=(e, ~\mu, ~\tau)$, to the mass eigenstates, 
i=(1, 2, 3), is called the Pontecorvo-Maki-Nakagawa-Sakata matrix 
(PMNS)\cite{PMNS},
that is 
\begin{equation}
|\nu_\alpha\rangle  =  U_{\alpha,i} ~|\nu_i \rangle
\end{equation}
With the above identification of the mass eigenstates the standard
PMNS matrix parametrization is given by\cite{majorana}
\begin{eqnarray}
\hspace*{-0.5cm}
&  U_{\alpha i}  =  & 
\label{mns}
\end{eqnarray}
\vspace{-0.3cm}
\begin{eqnarray}
&
\hspace{-0.35cm}
\left( \begin{array}{ccc}
  1       &    & \\
    & c_{23} &  s_{23} \\
  & - s_{23} & c_{13} 
\end{array} \right) \, 
%
\left( \begin{array}{ccc}
  c_{13} &  & s_{13} e^{-i\delta} \\
& 1 & \\
  - s_{13} e^{i\delta} & & c_{13} 
\end{array} \right) \, 
%
\left( \begin{array}{ccc}
  c_{12}       & s_{12}  & \\
- s_{12} & c_{12} & \\
& & 1 
\end{array} \right) \,
%
%
& \nonumber
\end{eqnarray}
\vspace{-0.35cm}
\begin{eqnarray}
& = & \nonumber \\[0.25cm]
& \hspace{-0.35cm}
\left( \begin{array}{ccc}
  c_{13} c_{12}       & c_{13} s_{12}  & s_{13} e^{-i\delta} \\
- c_{23} s_{12} - s_{13} s_{23} c_{12} e^{i\delta}
& c_{23} c_{12} - s_{13} s_{23} s_{12} e^{i\delta}
& c_{13} s_{23} \\
    s_{23} s_{12} - s_{13} c_{23} c_{12} e^{i\delta}
& - s_{23} c_{12} - s_{13} c_{23} s_{12} e^{i\delta}
& c_{13} c_{23} 
\end{array} \right) \, & \nonumber  
\end{eqnarray}
where $s_{ij}$ and $c_{ij}$ are shorthand for $\sin \theta_{ij}$ and 
$\cos \theta_{ij}$ respectively.
The factorized form of this PMNS mixing matrix, eqn(\ref{mns}),
is useful because it allows
an identification between the mixing angles of the three component
picture and those of the two component analysis reported 
by the experiments.
Thus,
\begin{equation}
\theta_{23} \cong \theta_{atm}, \quad
\theta_{12} \cong \theta_{sol},
\quad {\rm and} \quad \theta_{13} \cong \theta_{chz}.
\end{equation}
This is an excellent 
approximation because of the smallest
of the ratio of the solar to atmospheric 
$\delta m^2$, $\delta m^2_{sol}/\delta m^2_{atm}$,
{\it and} $\sin^2 \theta_{chz}$. 
A recent detailed 3 neutrino analysis 
can be found
in \cite{valle3}.
The phase factor $\delta$,
($0 \leq  \delta < 2\pi$),
doesn't occur in any two component analysis 
however this phase allows for the possibility of 
CP violation in three neutrino oscillations.

\begin{figure*}
\epsfig{file=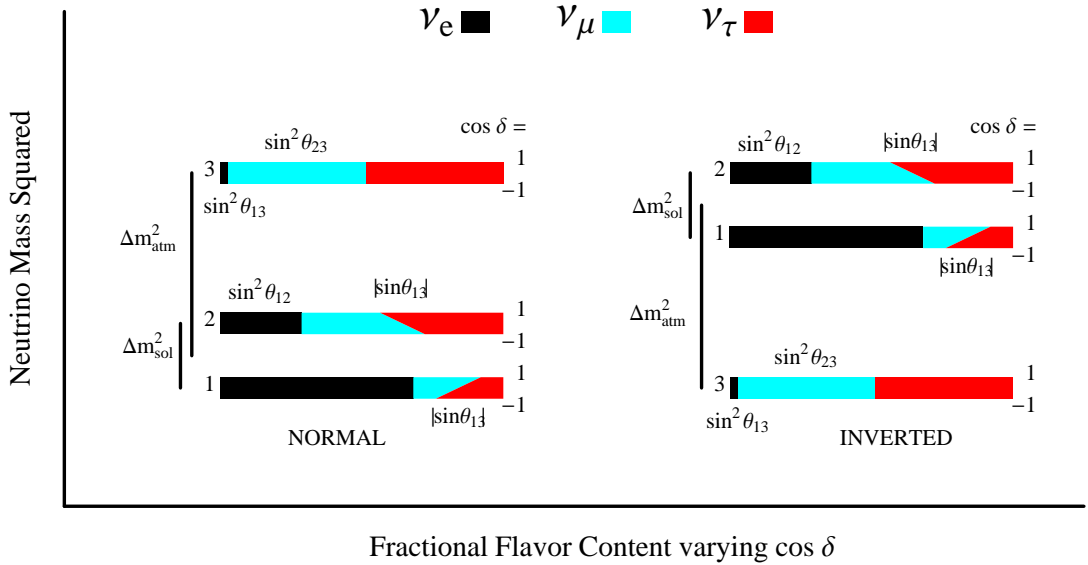, width=15cm} 
\caption{The range of probability of finding the $\alpha$-flavor in the i-th 
mass eigenstate as indicated as the CP-violating phase, $\delta$, is varied.
The bottom of the bars is for the minimum
allowed value of $\cos \delta = -1$ and the top of the bars
is for the maximum value of   $\cos \delta = 1$.
The other mixing parameters are held fixed:
$\sin^2 \theta_{12} = 0.30$, $\sin^2 \theta_{13} = 0.03$ and
$\sin^2 \theta_{23} = 0.50$. 
The maximum to minimum variation of the fractional flavor content
of $\mu$ or $\tau$ in mass eigenstates 1 or 2 is very close to 
$\sin \theta_{13}$.
The only parameter in the PMNS mixing matrix this figure is not sensitive to
is the sign of $\sin \delta$.
}
\label{fig2}
\end{figure*}

An extremely useful way to understand the meaning of the
various mixing angles
is to relate them to the probability of finding the 
$\alpha$-flavor in the i-th mass eigenstate.
This probability is given by the absolute square of the PMNS 
matrix elements, $|U_{\alpha i}|^2$.
Thus the probability of finding $\nu_e$ in the 3-th neutrino mass eigenstate
is just $\sin^2 \theta_{13}$  which is known from the Chooz data to be
no larger than a few per cent ($<$ 3\%). 
Similarly the probability of finding  
$\nu_\mu$ ($\nu_\tau$) in the 3-th mass eigenstate
is just $\cos^2 \theta_{13} \sin^2 \theta_{23} \approx \sin^2 \theta_{23}$
($\cos^2 \theta_{13} \cos^2 \theta_{23} \approx \cos^2 \theta_{23}$)
since $\cos^2 \theta_{13}$ is very close to unity.
Also the probability of finding the $\nu_e$ in the 2-th mass eigenstate
is just $\cos^2 \theta_{13} \sin^2 \theta_{12} \approx \sin^2 \theta_{12}$.
Since the $^8B$ solar neutrinos exit the sun as nearly a pure $\nu_2$
neutrino mass eigenstate, due to matter effects\cite{parke},
the measurement of the Charge Current to
Neutral Current (CC/NC) ratio by SNO is a direct measurement of 
$\sin^2 \theta_{12}$ up to small corrections.

In general the probability of finding the 
$\alpha$-flavor in the i-th mass eigenstate, P$_\nu$($\alpha$, i)
is given by
\begin{eqnarray}
\hspace*{-0.5cm}
& &  \quad \quad P_\nu (\alpha, i) = |U_{\alpha i}|^2 \approx 
\label{mns_sq} \\[0.1in]
& &
\left( \begin{array}{ccc}
  c^2_{12}       &  s^2_{12}  & s^2_{13} \\
c^2_{23} s^2_{12} + K s_{13} \cos{\delta}  \quad
&  c^2_{23} c^2_{12} - K s_{13}\cos{\delta} \quad
&  s^2_{23} \\
    s^2_{23} s^2_{12} - K s_{13} \cos{\delta} \quad
& s^2_{23} c^2_{12} + K s_{13} \cos{\delta} \quad
& c^2_{23} 
\end{array} \right) \, \nonumber
\end{eqnarray}
where $K = \frac{1}{2} \sin{2\theta_{12}} \sin{2\theta_{23}}
~(\approx \frac{1}{2})$
and terms of order $\sin^2 \theta_{13}$ have been dropped except in
the (e,3) component which otherwise would be zero.
Note, that up to this order the sum of each row and each column
of this probability matrix adds up to one as required by unitarity.
The probabilities ($\mu$,1), ($\mu$,2), ($\tau$,1) and ($\tau$,2)
all depend linearly on $\sin \theta_{13} \cos \delta$ whose sign 
is determined by $\cos \delta$ and the magnitude can be quite significant
compared to the terms independent of $\sin \theta_{13}$ in these 
probabilities.

Translating the mixing angle information reported by the experiments
into ranges of probability of finding the $\alpha$-flavor in 
the i-th mass eigenstate we obtain
\begin{eqnarray}
0.25  < & \sin^2 \theta_{12} \cong P_\nu(e,2)   &<  0.33 \nonumber \\
0.35  < & \quad \sin^2 \theta_{23} \cong P_\nu(\mu,3) \quad &<  0.65 
\label{prob_limits} \\
         & \sin^2 \theta_{13} \equiv P_\nu(e,3)   &<  0.03 \nonumber
\end{eqnarray}
at the 90\% confidence level.
Clearly, using the probability metric, i.e. $\sin^2 \theta$,
our current information of the solar mixing is significantly
better than that of the atmospheric mixing.
This occurs because $\sin^2 2 \theta$ is a poor measure of $\sin^2 \theta$ 
near $\sin^2 \theta = \frac{1}{2}$.
Eqn(\ref{mns_sq}) and (\ref{prob_limits}) can be used to calculate 
the ranges for all the
other the probabilities with the unknown $\cos \delta$ varying from
-1 to +1.  

In the past, the central value of all of these probabilities
has been presented in a bar graph with a separate horizontal bar for each
neutrino mass eigenstate with color and/or shading coding for each
of the neutrino flavors.
This is a very useful pictorial way of presenting all of the neutrino
mixing data with a physical interpretation.
In this letter we extend this diagram to include the range of possible
probabilities allowed by the data. 
To do this we make use of the thickness of the bars so that the bottom
of the bar represents the minimum allowed value for a particular
parameter
and the top of the bar represents the maximum
allowed value of this parameter.
With these figures one can easily see how the range of parameters
effects the flavor probabilities for any given mass eigenstate.
Remember that these flavor probabilities are the absolute squares of
the elements of the PMNS mixing matrix.
Below we discuss a few important examples of such figures.

In Fig.~\ref{fig1} we have held all mixing parameters
fixed (the values are given in the caption),
except $\sin^2 \theta_{23}$ which varies 
through out it's allowed range.
Clearly the probability of finding the
$\nu_\mu$ or $\nu_\tau$ flavors in all of the neutrino mass eigenstates
varies significantly because of the uncertainty in $\sin^2 \theta_{23}$.
This uncertainty in  $\sin^2 \theta_{23}$ is of particular importance
for the proposed Long Baseline neutrino experiments
searching for $\nu_\mu \rightarrow \nu_e$, 
JParc to SK\cite{JHF2SK} and NuMI-Off-Axis\cite{NOA},
as the leading term in the oscillation probability
is proportional to
$\sin^2 \theta_{23} \sin^2 \theta_{13}$.
 
Similarly, in Fig.~\ref{fig2} we have varied the CP violating phase
so that $\cos \delta$ varies from -1 to +1 keep the other angles fixed.  
Notice that for values of $\sin^2 \theta_{13}$ close to the maximum 
allowed by the Chooz experiment the probabilities   
($\mu$,1), ($\mu$,2), ($\tau$,1) and ($\tau$,2) have significant 
dependence on the CP violating phase $\delta$ through the variable
$\cos \delta$.  
The range of variation of any of these probabilities is linear in
$\sin \theta_{13}$ so that  $\sin^2 \theta_{13}$
must be very small before this range is insignificant.
Determination of the phase $\delta$ will most likely come from 
future Long Baseline experiments as the asymmetry between
$\nu_\mu \rightarrow \nu_e$ 
and $\bar{\nu}_\mu \rightarrow \bar{\nu}_e$, which can approach one,
is directly proportional to $\sin \delta$ if the energy and baseline
are tuned to a peak in the oscillations.

Other variations of this figure are possible\cite{web} 
including the possibility
of varying more than one variable at a time e.g. one could choose the 
bottom of the bars so that say P$_\nu$($\mu$,1) is minimum 
and the top such that
P$_\nu$($\mu$,1) is maximum for the allowed ranges of all 
the mixing angles.
Another possibility is to use the maximum range of each probability without
including the correlations between these probabilities.

In summary we have argued that using $\sin^2 \theta$ instead of 
the numerous other ways that exist in the literature allows
for a unified treatment of the neutrino mixing parameters
which has direct physical interpretation.

\begin{itemize}
\item $\sin^2 \theta_{sol}$ is the probability of finding the
electron neutrino flavor in the heavier of the solar
doublet of neutrino mass eigenstates (labeled 2) up to
small corrections
and is directly measured by the SNO
Charge Current to Neutral Current ratio. 
\item $\sin^2 \theta_{atm}$ is the probability of finding the
muon neutrino flavor in the isolated neutrino mass eigenstate
(labeled 3) up to small corrections.
\item $\sin^2 \theta_{chz}$ is the probability of finding the
electron neutrino flavor in the isolated neutrino mass eigenstate 
(labeled 3).
\end{itemize}
We have extended the usual pictorial presentation of the 
absolute square of the PMNS matrix elements, i.e. the probabilities of finding 
a given flavor in a particular mass eigenstate, so as to incorporate
the uncertainties in our knowledge of the mixing parameters.
Together, these two features provide a coherent, unified, graphical summary
of our present knowledge of the neutrino sector.

\bigskip
We thank all the organizers of WIN'03 which provided the inspiration for 
the figures that lead to this paper and to Peter Cooper for a careful
reading and suggestions on the manuscript.
Fermilab is operated under DOE contract DE-AC02-76CH03000.



\begin{thebibliography}{99}
\bibitem{majorana}
For Majorana neutrinos $U$ is replaced by 
$U\,{\rm diag}[1,e^{i\alpha_2},e^{i\alpha_3}]$.
Since neutrino oscillations depend on $U_{\alpha i}U^*_{\beta i}$
these Majorana phases, $(\alpha_2, \alpha_3)$, are unobservable in 
neutrino oscillation but are important for neutrino-less double
$\beta$-decay.
\end{thebibliography}
\end{document}